\documentclass[aps,prl,twocolumn,floatfix,nofootinbib,showpacs,superscriptaddress]{revtex4-1}

\usepackage{amsmath,amsfonts,amssymb,bm}
\usepackage{dsfont}
\usepackage{graphicx}
\usepackage{color}
\usepackage{pzccal}
\usepackage{soul}
\definecolor{purple}{rgb}{0.5,0,0.5}
\definecolor{blue}{rgb}{0.0,0,0.9}

\begin{document}

\title{Pion distribution amplitude from lattice-QCD}

\author{I.\,C.~Clo\"et}
\affiliation{Physics Division, Argonne National Laboratory, Argonne, Illinois 60439, USA}

\author{L.~Chang}
\affiliation{Institut f\"ur Kernphysik, Forschungszentrum J\"ulich, D-52425 J\"ulich, Germany}

\author{C.\,D.~Roberts}
\affiliation{Physics Division, Argonne National Laboratory, Argonne, Illinois 60439, USA}

\author{S.\,M.~Schmidt}
\affiliation{Institute for Advanced Simulation, Forschungszentrum J\"ulich and JARA, D-52425 J\"ulich, Germany}

\author{P.\,C.~Tandy} \affiliation{Center for Nuclear Research, Department of
Physics, Kent State University, Kent OH 44242, USA}

\date{28 May 2013}

\begin{abstract}
A method is explained through which a pointwise accurate approximation to the pion's valence-quark distribution amplitude (PDA) may be obtained from a limited number of moments. In connection with the single nontrivial moment accessible in contemporary simulations of lattice-regularised quantum chromodynamics (QCD), the method yields a PDA that is a broad concave function whose pointwise form agrees with that predicted by Dyson-Schwinger equation analyses of the pion.  Under leading-order evolution, the PDA remains broad to energy scales in excess of $100\,$GeV, a feature which signals persistence of the influence of dynamical chiral symmetry breaking.  Consequently, the asymptotic distribution, $\varphi_\pi^{\rm asy}(x)$, is a poor approximation to the pion's PDA at all such scales that are either currently accessible or foreseeable in experiments on pion elastic and transition form factors.  Thus, related expectations based on $\varphi_\pi^{\rm asy}(x)$ should be revised.
\end{abstract}


\maketitle

%
The light-front wave-function of an interacting quantum system, $\varphi(x)$, provides a connection between dynamical properties of the underlying relativistic quantum field theory and notions familiar from nonrelativistic quantum mechanics.  In particular, although particle number conservation is generally lost in relativistic quantum field theory, $\varphi(x)$ has a probability interpretation.  It can therefore translate features that arise purely through the infinitely-many-body nature of relativistic quantum field theory into images whose interpretation seems more straightforward \cite{Keister:1991sb,Brodsky:1997de,Chang:2013pq}.

With $\varphi(x)$ in hand, the impact of phenomena that are essentially quantum field theoretical in origin may be expressed via wave-function overlaps.
Such overlaps are familiar in all disciplines and associated with a long-established interpretation.  For example, the (leading-twist) wave-function of a meson is an amplitude that describes the momentum distribution of a quark and antiquark in the bound-state's simplest (valence) Fock state.  The amplitude, $\varphi(x)$, is a process-independent expression of intrinsic properties of the composite system.

Seemingly, the simplest composite systems in nuclear and particle physics are the pions.  This isospin triplet of (unusually) low-mass states are constructed from valence $u$- and $d$-quarks.
As a process-independent expression of pion properties, $\varphi_\pi(x)$ is a crucial element in computing the leading-twist and leading-order in $\alpha$-strong results for pion elastic and transition form factors \cite{Farrar:1979aw,Efremov:1979qk,Lepage:1980fj}.  For many years, predictions obtained with such formulae have served as motivation for crucial experiments designed to test QCD; e.g., Refs.\,\cite{Volmer:2000ek,Horn:2006tm,Huber:2008id,Behrend:1990sr,Gronberg:1997fj,
Aubert:2009mc,Uehara:2012ag}.

Regarding the pion, however, appearances have long been deceiving.
The unusually low mass of these states signals the intimate connection between dynamical chiral symmetry breaking (DCSB) and the existence and properties of pions.
This connection is fascinating because DCSB is a striking emergent feature of QCD, which plays a critical role in forming the bulk of the visible mass in the Universe \cite{national2012Nuclear} and is expressed in numerous aspects of the spectrum and interactions of hadrons; e.g., the large splitting between parity partners \cite{Chang:2011ei,Chen:2012qr} and the existence and location of a zero in some hadron form factors \cite{Wilson:2011aa,Cloet:2013gva}.
%
As emphasised by the successful application of chiral perturbation theory at soft scales, an explanation of pion properties is only possible within an architecture that faithfully represents chiral symmetry and the pattern by which it is broken in QCD.  In order to chart the pion's internal structure, one must unify this with a direct connection to the parton dynamics of QCD.  The Dyson-Schwinger equation (DSE) framework \cite{Chang:2011vu,Bashir:2012fs} effects such a union in the continuum and, with recent algorithmic advances, it may now be employed for the computation of light-front quantities such as $\varphi_\pi(x)$ \cite{Chang:2013pq}.

Hitherto, we have not explained the argument $x$, upon which the pion's valence-quark PDA depends.  This variable expresses the light-front fraction of
the pion's total-momentum carried by the valence quark, which is equivalent to the momentum fraction carried by the valence-quark in the infinite-momentum frame.  Momentum conservation entails that the valence antiquark carries $(1-x)$.  Since the neutral pion is an eigenstate of the charge conjugation operator, $\varphi_\pi(x)=\varphi_\pi(1-x)$.

We have, in addition, omitted an argument that is crucial in understanding and employing $\varphi_\pi(x)$.  Namely, the PDA is also a function of the momentum-scale $\zeta$ or, equivalently, the length-scale $\tau=1/\zeta$, which characterises the process in which the pion is involved.  On the domain within which QCD perturbation theory is valid, the equation describing the $\tau$-evolution of $\varphi_\pi(x;\tau)$ is known and has the solution  \cite{Efremov:1979qk,Lepage:1980fj}
\begin{eqnarray}
\label{PDAG3on2}
\varphi_\pi(x;\tau) &=& \varphi^{\rm asy}_\pi(x)
\bigg[ 1 + \!\! \sum_{j=2,4,\ldots}^{\infty} \!\! \!\! a_j^{3/2}(\tau) \,C_j^{(3/2)}(2 x -1) \bigg],\;\;\\
\varphi^{\rm asy}_\pi(x) &=& 6 x (1-x)\,, \label{phiasy}
\end{eqnarray}
where $\{C_j^{(3/2)},j=1,\ldots,\infty\}$ are Gegenbauer polynomials of order $\alpha=3/2$ and the expansion coefficients $\{a_j^{3/2},j=1,\ldots,\infty\}$ evolve logarithmically with $\tau$, vanishing as $\tau\to 0$.  (These features owe to the fact that in the neighbourhood $\tau \Lambda_{\rm QCD} \simeq 0$, where $\Lambda_{\rm QCD}\sim 0.2\,$GeV, QCD is invariant under the collinear conformal group
SL$(2;\mathbb{R})$
\cite{Brodsky:1980ny,Braun:2003rp}.  Indeed, the Gegenbauer-$\alpha=3/2$ polynomials are merely irreducible representations of this group.  A correspondence with the spherical harmonics expansion of the wave-functions for $O(3)$-invariant systems in quantum mechanics is plain.)

In the absence of additional information, it has commonly been assumed that at any length-scale $\tau$, a useful approximation to $\varphi_\pi(x;\tau)$ is obtained by using just the first few terms of the expansion in Eq.\,\eqref{PDAG3on2}.  (This assumption has led to models for $\varphi_\pi(x)$ whose pointwise behaviour is not concave on $x\in[0,1]$; e.g., to ``humped'' distributions \cite{Chernyak:1983ej}.)  Whilst the assumption is satisfied on $\tau \Lambda_{\rm QCD} \simeq 0$, it is hard to justify at the length-scales available in typical contemporary experiments, which correspond to $\zeta \simeq 2\,$GeV.  This is emphasised by the fact that within the domain $\tau \Lambda_{\rm QCD} \simeq 0$, the pion's valence-quark parton distribution function $u_{\rm v}^\pi(x) \approx \delta(x)$, which is far from valid at currently accessible scales \cite{Hecht:2000xa,Aicher:2010cb,Nguyen:2011jy}.

To illustrate these remarks, consider that a value
\begin{equation}
\label{a2Old}
a_2^{3/2}(\tau_2)=0.201(114)\,,
\end{equation}
$\tau_2=1/[2\,{\rm GeV}]$, was obtained using Eq.\,\eqref{PDAG3on2} as a tool for expressing the result of a numerical simulation of lattice-regularised QCD \cite{Braun:2006dg}.  This indicates a large correction to the asymptotic form, $\varphi^{\rm asy}_\pi(x)$, and gives no reason to expect that the ratio $a_4^{3/2}(\tau_2)/a_2^{3/2}(\tau_2)$ is small.  Now, at leading-logarithmic accuracy, the moments in Eq.\,\eqref{PDAG3on2} evolve from $\tau_2\to \tau$ as follows \cite{Efremov:1979qk,Lepage:1980fj}:
\begin{equation}
\label{llevolution}
a_j^{3/2}(\tau) = a_j^{3/2}(\tau_2) \left[\frac{\alpha_s(\tau_2)}{\alpha_s(\tau)}\right]^{\gamma_n^{(0)}/\beta_0},
\end{equation}
where the one-loop strong running-coupling is
\begin{equation}
\alpha_s(\tau) = \frac{2 \pi}{\beta_0\,\ln(1/[\tau \Lambda_{\rm QCD}])},
\end{equation}
with $\beta_0 = 11 - (2/3) n_f$, and
\begin{equation}
\gamma_n^{(0)} = C_F\left[3 + \frac{2}{(j+1)\,(j+2)} - 4\,\sum_{k=1}^{j+1}\,\frac{1}{k}\right],
%
\end{equation}
where $C_F = 4/3$ and $n_f$ is the number of active flavours.  Using $n_f=4$ and $\Lambda_{\rm QCD}=0.234\,$GeV for illustration \cite{Qin:2011dd}, it is necessary to evolve to $\tau_{100}=1/[100\,{\rm GeV}]$, before $a_2^{3/2}(\tau)$ even falls to 50\% of its value in Eq.\,\eqref{a2Old}.  The $a_4^{3/2}$ coefficient still holds 37\% of its value at $\tau_{100}$.  This pattern is qualitatively preserved with higher order evolution \cite{Mikhailov:1984ii,Mueller:1998fv}.  These observations suggest that the asymptotic domain lies at very large momenta indeed.  


%
As observed already, the pion's valence-quark PDA was recently computed using QCD's Dyson-Schwinger equations (DSEs) \cite{Chang:2013pq}.  At the scale $\zeta=2\,$GeV, $\varphi_\pi(x;\tau_2)$ is much broader than the asymptotic form, $\varphi^{\rm asy}_\pi(x)$ in Eq.\,\eqref{phiasy}.  Indeed, the power-law dependence is better characterised by $x^{\alpha_-} (1-x)^{\alpha_-}$ with $\alpha_- \approx 0.3$, a value very different from that associated with the asymptotic form; viz., $\alpha_-^{\rm asy}=1$.  Importantly, this dilation is a long-sought and unambiguous expression of dynamical chiral symmetry breaking (DCSB) on the light-front \cite{Brodsky:2010xf,Chang:2011mu,Brodsky:2012ku}.

If one insists on using Eq.\,\eqref{PDAG3on2} to represent such a broad distribution, then $a_{14}^{3/2}$ is the first expansion coefficient whose magnitude is less-than 10\% of $a_{2}^{3/2}$.  For the following reasons, we do not find this surprising.  The polynomials $\{C_j^{(3/2)}(2x-1),j=1,\ldots,\infty\}$ are a complete orthonormal set on $x\in[0,1]$ with respect to the measure $x(1-x)$.  Just as any attempt to represent a box-like curve via a Fourier series will inevitably lead to slow convergence and spurious oscillations, so does the use of Gegenbauer polynomials of order $\alpha=3/2$ to represent a function better matched to the measure $x^{0.3} (1-x)^{0.3}$.  This latter measure is actually associated with Gegenbauer polynomials of order $\alpha = 4/5$.  Observations such as these led to the method adopted in Ref.\,\cite{Chang:2013pq}.

As a framework within continuum quantum field theory, the DSE study of Ref.\,\cite{Chang:2013pq} was able to reliably compute arbitrarily many moments of the PDA, using
\begin{equation}
f_\pi (n\cdot P)^{m+1} \langle x^m\rangle =
{\rm tr}_{\rm CD}
Z_2 \! \int_{dq}^\Lambda \!\!
(n\cdot q_\eta)^m \,\gamma_5\gamma\cdot n\, \chi_\pi(q;P)\,,
\label{phimom}
\end{equation}
where: $f_\pi$ is the pion's leptonic decay constant; the trace is over colour and spinor indices; $\int_{dq}^\Lambda$ is a Poincar\'e-invariant regularisation of the four-dimensional integral, with $\Lambda$ the ultraviolet regularization mass-scale; $Z_{2}(\zeta,\Lambda)$ is the quark wave-function renormalisation constant, with $\zeta$ the renormalisation scale; $n$ is a light-like four-vector, $n^2=0$; $P$ is the pion's four-momentum; and $\chi_\pi$ is the pion's Bethe-Salpeter wave-function
\begin{equation}
\chi_\pi(q;P) = S(q_\eta) \Gamma_\pi(q;P) S(q_{\bar \eta})\,,
\label{chipi}
\end{equation}
with $\Gamma_\pi(q;P)$ the Bethe-Salpeter amplitude,  $S$ the dressed light-quark propagator, and $q_\eta = q + \eta P$, $q_{\bar\eta} = q - (1-\eta) P$, $\eta\in [0,1]$.  Owing to Poincar\'e covariance, no observable can legitimately depend on $\eta$.

In order to inform expectations about the nature of the PDA that is reconstructed from the moments in Eq.\,\eqref{phimom}, we repeat that the pion multiplet contains a charge-conjugation eigenstate.  Therefore, the peak in the leading Chebyshev moment of each of the three significant scalar functions that appear in the expression for $\Gamma_\pi(q;P)$ occurs at $2 k_{\rm rel}:=q_\eta + q_{\bar\eta} = 0$; i.e., at zero relative momentum \cite{Maris:1997tm,Qin:2011xq}.  Moreover, these Chebyshev moments are monotonically decreasing with $k_{\rm rel}^2$.  Such observations suggest that $\varphi_\pi(x)$ should exhibit a single maximum, which appears at $x=1/2$; i.e.,  $\varphi_\pi(x)$ is a symmetric, concave function on $x\in [0,1]$.

In Ref.\,\cite{Chang:2013pq}, from fifty moments produced by Eq.\,\eqref{phimom}, the PDA was reconstructed using Gegenbauer polynomials of order $\alpha$, with this order -- the value of $\alpha$ -- determined by the moments themselves, not fixed beforehand.  Namely, with
\begin{equation}
\label{PDAGalpha}
\varphi_\pi(x;\tau) = N_\alpha\, x^{\alpha_-} (1-x)^{\alpha_-}
\bigg[ 1 + \sum_{j=2,4,\ldots}^{j_s} \!\!\!\! a_j^{\alpha}(\tau)\, C_j^{(\alpha)}(2 x -1) \bigg],
\end{equation}
where $\alpha_-=\alpha-1/2$ and $N_\alpha = \Gamma(2\alpha+1)/[\Gamma(\alpha+1/2)]^2$, very rapid progress from the moments to a converged representation of the PDA was obtained.  Indeed, $j_s=2$ was sufficient, with $j_s=4$ producing no change in a plotted curve that was greater than the line-width.  Naturally, once obtained in this way, one may project $\varphi_\pi(x;\tau)$ onto the form in Eq.\,\eqref{PDAG3on2}; viz., for $j=2,4,\ldots\,$,
\begin{equation}
\label{projection}
a_j^{3/2} = \frac{2}{3}\ \frac{2\,j+3}{(j+2)\,(j+1)}\int_0^1 dx\, C_j^{(3/2)}(2\,x-1)\,\varphi_\pi(x),
\end{equation}
therewith obtaining all coefficients necessary to represent any computed distribution in the conformal form without ambiguity or difficulty.

We advocate taking this approach a step further; viz., adopting it, too, when one is presented even with only limited information on $\varphi_\pi(x;\tau)$.  In this connection, consider that since discretised spacetime does not possess the full rotational symmetries of the Euclidean continuum, then, with current algorithms, only one nontrivial moment of $\varphi_\pi(x)$ can be computed using numerical simulations of lattice-regularised QCD.  Thus, in Ref.\,\cite{Braun:2006dg}, using two flavors of dynamical, $O(a)$-improved Wilson fermions and linearly extrapolating to the empirical pion mass, $\hat m_\pi$, from results at $m_\pi^2/\hat m_\pi^2 = 20,35,50$, the following lone result for the pion is found:
\begin{equation}
\label{latticemoment}
\int_0^1 dx\, (2 x - 1)^2 \, \varphi_\pi(x,\tau_2) = 0.27\pm 0.04\,.
\end{equation}

This single moment can only produce one piece of information about $\varphi_\pi(x;\tau)$; and, as described in connection with Eq.\,\eqref{a2Old}, it was used in Ref.\,\cite{Braun:2006dg} to constrain $a_2^{3/2}(\tau_2)$ in Eq.\,\eqref{PDAG3on2} and therewith produce a ``double-humped'' PDA.  Notably, following Ref.\,\cite{Chang:2013pq}, it is straightforward to establish that a double-humped form lies within the class of distributions produced by a pion Bethe-Salpeter amplitude that may be characterised as vanishing at zero relative momentum, instead of peaking thereat.

\begin{figure}[t]
\begin{centering}
\includegraphics[clip,width=0.90\linewidth]{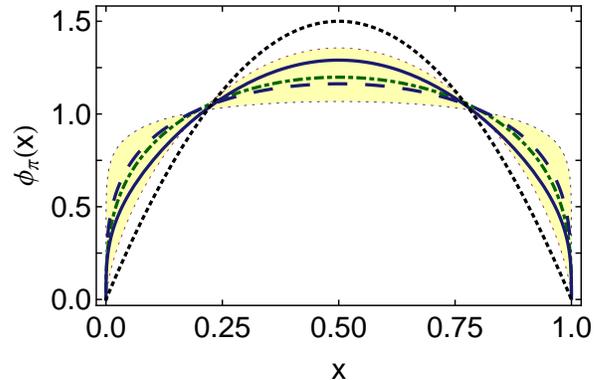}
\end{centering}
\caption{\label{Fig1}
\emph{Dot-dashed curve}, embedded in the shaded region, $\varphi_\pi(x;\tau_2)$ in Eq.\,\protect\eqref{pionPDAlattice}.  The shaded region indicates the extremes allowed by the errors on $\alpha_-$.
For comparison, the DSE results obtained in Ref.\,\protect\cite{Chang:2013pq} are also depicted: \emph{solid curve}, $\varphi_\pi(x;\tau_2)$ obtained with the best DSE truncation currently available, which includes important features of DCSB in building the kernels; and \emph{dashed curve}, result obtained in rainbow-ladder truncation.
The \emph{dotted curve} is $\varphi_\pi^{\rm asy}(x)$.}
\end{figure}


%
Now, suppose instead that one analyses the single unit of information in Eq.\,\eqref{latticemoment} using Eq.\,\eqref{PDAGalpha} but discarding the sum, a procedure which acknowledges implicitly that the pion's PDA should exhibit a single maximum at $x=1/2$.  Then, Eq.\,\eqref{latticemoment} constrains $\alpha$, with the result
\begin{equation}
\label{pionPDAlattice}
\varphi_\pi(x;\tau_2) = N_\alpha \, x^{\alpha_-} (1-x)^{\alpha_-},\;
\alpha_- = 0.35^{+ 0.32 = 0.67}_{-0.24=0.11},
\end{equation}
which is depicted in Fig.\,\ref{Fig1}.  Employed thus, the lattice-QCD result, Eq.\,\eqref{pionPDAlattice}, produces a concave amplitude in agreement with contemporary DSE studies and confirms that the asymptotic distribution, $\varphi_\pi^{\rm asy}(x)$, is not a good approximation to the pion's PDA at $\zeta=2\,$GeV.

Equation~\eqref{pionPDAlattice} actually favours the DSE result obtained with the interaction of Ref.\,\cite{Qin:2011dd} and the rainbow-ladder (RL) truncation.  This truncation is the leading order in a systematic, symmetry-preserving scheme \cite{Munczek:1994zz,Bender:1996bb} that has widely been used with success in explaining properties of ground-state pseudoscalar and vector mesons \cite{Maris:2003vk} and the nucleon and $\Delta$ \cite{Eichmann:2011ej,Segovia:2013rca}.
The other DSE curve was obtained with the same interaction but using novel representations of the gap and Bethe-Salpeter kernels that incorporate important, essentially nonperturbative features of DCSB, which it is impossible to recover in RL truncation or any stepwise improvement thereof \cite{Chang:2009zb,Chang:2010hb,Chang:2011ei}.  The solid curve should therefore provide the more realistic result.  That the PDA inferred from Eq.\,\eqref{latticemoment} is closer to the RL result is nonetheless readily understood.  As just described, RL computations omit important features of DCSB and, in being obtained by linearly extrapolating from large pion masses, so, effectively, does the lattice result.  We anticipate that improved lattice simulations will produce a PDA in better agreement with the solid curve in Fig.\,\ref{Fig1}.

To illustrate and emphasise that information is gained using the procedure we advocate but not lost, we list the first three Gegenbauer-$\alpha=3/2$ moments computed by reprojecting Eq.\,\eqref{pionPDAlattice} onto the expansion in Eq.\,\eqref{PDAG3on2}, using Eq.\,\eqref{projection}:
\begin{subequations}
{\allowdisplaybreaks
\begin{eqnarray}
\label{a2New}
a_2^{3/2}(\tau_2) & = & 0.20 \pm 0.12 \,,\\
\label{a4New} a_4^{3/2}(\tau_2) & = & 0.093 \pm 0.064 \,,\\
\label{a6New} a_6^{3/2}(\tau_2) & = & 0.055 \pm 0.041 \,.
\end{eqnarray}}
\end{subequations}
\hspace*{-0.4em}Naturally, the result in Eq.\,\eqref{a2New} is equivalent to that in Eq.\,\eqref{a2Old}, and Eqs.\,\eqref{a4New}, \eqref{a6New} provide new information, which might either be checked by, or used to inform, other approaches to the problem of computing $\varphi_\pi(x)$; e.g., Refs.\,\cite{Frederico:1994dx,Brodsky:2006uqa,Radyushkin:2009zg,Agaev:2010aq,Bakulev:2012nh,
ElBennich:2012ij}.  Moreover, with Eq.\,\eqref{pionPDAlattice} one obtains
\begin{equation}
\varphi_\pi(x=1/2;\tau_2) = 1.20^{+0.16=1.36}_{-0.13=1.07}\,,
\end{equation}
which agrees with the result $\varphi_\pi(1/2) = 1.2 \pm 0.3$ obtained using QCD sum rules \cite{Braun:1988qv}.

\begin{figure}[t]
\begin{centering}
\includegraphics[clip,width=0.90\linewidth]{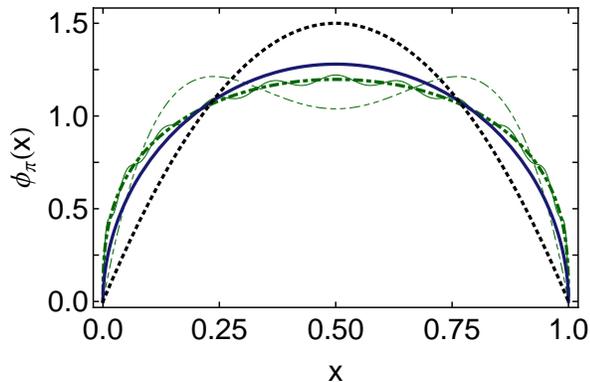}
\end{centering}
\caption{\label{Fig2}
\emph{Dot-dashed curve}, $\varphi_\pi(x;\tau_2)$ in Eq.\,\protect\eqref{pionPDAlattice}; oscillatory \emph{thin solid curve}, Gegenbauer-$\alpha=3/2$ representation obtained with $10$ nontrivial moments ($a_{20}^{3/2}/a_2^{3/2}=0.044$); and \emph{thin dot-dot-dashed curve}, Gegenbauer-$\alpha=3/2$ representation obtained with just $1$ nontrivial moment ($a_{2}^{3/2}=0.20$, Eq.\,\protect\eqref{a2New}.).
\emph{Solid curve}, $\varphi_\pi(x;\tau_{10})$ in Eq.\,\protect\eqref{pionPDAlattice10}; i.e., leading-order evolution of $\varphi_\pi(x;\tau_2)$ to $\tau_{10}=1/[10\,{\rm GeV}]$, which corresponds to a hard scale of $100\,$GeV$^2$.
The dotted curve is $\varphi_\pi^{\rm asy}(x)$.}
\end{figure}

As noted above, one may accurately compute arbitrarily many Gegenbauer-$\alpha=3/2$ moments by reprojecting the result in Eq.\,\eqref{pionPDAlattice} onto the Gegenbauer-$\alpha=3/2$ basis, Eqs.\,\eqref{PDAG3on2}, \eqref{projection}.  It is therefore straightforward to evolve Eq.\,\eqref{pionPDAlattice} to any scale $\zeta$ that might be necessary in order to consider a given process.  This is illustrated in Fig.\,\ref{Fig2}.
To prepare the figure, we expressed Eq.\,\eqref{pionPDAlattice} in the form of Eq.\,\eqref{PDAG3on2} with ten nontrivial moments, $\{a_j^{3/2}(\tau_2),j=2,\ldots,20\}$.
(N.B.\ The double-humped dot-dot-dashed curve, which depicts the result obtained if just the first moment is kept, highlights the limitation inherent in using Eq.\,\eqref{PDAG3on2} with limited information.)
Using the ten-moment expression and the leading-logarithmic formula, Eq.\,\eqref{llevolution}, those moments were evolved from $\zeta=2\,$GeV to $\zeta=10\,$GeV, producing a ten-moment representation of $\varphi_\pi(x;\tau_{10})$.  It, too, oscillates about a concave curve.  Working with the errors indicated in Eq.\,\eqref{pionPDAlattice}, one finds
\begin{equation}
\label{pionPDAlattice10}
\varphi_\pi(x;\tau_{10}) = N_\alpha \, x^{\alpha_-} (1-x)^{\alpha_-},\;
\alpha_- = 0.51^{+ 0.25 = 0.76}_{-0.20=0.31} .
\end{equation}
The ``central'' value of $\alpha_-=0.51$ is used to plot the thick, solid curve in Fig.\,\ref{Fig2}.  Using Eqs.\,\eqref{a2New}, \eqref{a4New} and the comment after Eq.\,\eqref{a2Old}, one finds that it is only for $\zeta\gtrsim 100\,$GeV that $a_2^{3/2}\lesssim 10$\% and $a_4^{3/2}/a_2^{3/2}\lesssim 30$\%.  Evidently, the influence of DCSB, which is the origin of the amplitude's breadth, persists to remarkably small length-scales.

Such calculations expose a critical internal inconsistency in Ref.\,\cite{Aitala:2000hb}, which claims to represent a direct measurement of $\varphi_\pi^2(x)$.  Using the reasoning therein, the two panels in Fig.\,3 correspond to $\zeta \approx 2\,$GeV (left) and $\zeta \approx 3\,$GeV (right).  The left panel depicts a broad distribution, for which Eq.\,\eqref{projection} yields $a_2^{3/2} \approx 0.27$, whereas the right panel is the asymptotic distribution, for which $a_2^{3/2}=0$; and, as illustrated by the material presented herein, it is impossible for QCD evolution from $\zeta=2 \to 3\,$GeV to connect these two curves.  Therefore, they cannot represent the same pion property and it is not credible to assert that $\varphi_\pi(x)$ is well represented by the asymptotic distribution for $\zeta^2 \gtrsim 10\,$GeV$^2$.  The assumptions which underly the claims in Ref.\,\cite{Aitala:2000hb} should be carefully re-examined.


%
The analysis presented herein establishes that 
contemporary DSE- and lattice-QCD computations, at the same scale, agree on the pointwise form of the pion's PDA, $\varphi_\pi(x;\tau)$.  This unification of DSE- and lattice-QCD results expresses a deeper equivalence between them, expressed, in particular, via the common behaviour they predict for the dressed-quark mass-function \cite{Bhagwat:2003vw,Bowman:2005vx,Bhagwat:2006tu,Roberts:2007ji}, which is a definitive signature of dynamical chiral symmetry breaking and the origin of the distribution amplitude's dilation.

Furthermore, the associated discussion supports a view that $\varphi_\pi^{\rm asy}(x)$ is a poor approximation to $\varphi_\pi(x;\tau)$ at all momentum-transfer scales that are either now accessible to experiments involving pion elastic or transition processes, or will become so in the foreseeable future \cite{Huber:2008id,Uehara:2012ag,Holt:2012gg,Dudek:2012vr}.  Available information indicates that the pion's PDA is significantly broader at these scales; and hence that predictions of leading-order, leading-twist formulae involving $\varphi^{\rm asy}_\pi(x)$ are a misleading guide to interpreting and understanding contemporary experiments.  At accessible energy scales a better guide is obtained by using the broad PDA described herein in such formulae.  This might be adequate for the charged pion's elastic form factor.  However, it will probably be necessary to consider higher twist and higher-order $\alpha$-strong corrections in controversial cases such as the $\gamma^\ast \gamma \to \pi^0$ transition form factor \cite{Roberts:2010rn,Agaev:2010aq,Brodsky:2011yv,Bakulev:2012nh}.


%
We thank S.\,J.~Brodsky and R.\,J.~Holt for valuable comments.
Work supported by:
Department of Energy, Office of Nuclear Physics, contract no.~DE-AC02-06CH11357;
For\-schungs\-zentrum J\"ulich GmbH;
and 
National Science Foundation, grant no.\
NSF-PHY1206187.



\end{document}